\documentclass{article}

\usepackage{microtype}
\usepackage{graphicx}
\usepackage{subfigure}
\usepackage{subcaption}
\usepackage{booktabs} 
\usepackage{float} 
\usepackage{listings,float} 
\usepackage{multirow} 
\floatstyle{ruled}
\newfloat{listing}{htbp}{lst}[section]
\floatname{listing}{Listing}
\usepackage{enumitem} 

\usepackage{hyperref}
\usepackage{hyperref}

\usepackage{tabularx,booktabs}   
\newcolumntype{L}[1]{>{\raggedright\arraybackslash}p{#1}}

\newcommand{\gc}{GUARDIAN-FC }

\usepackage[accepted]{icml2025}

\usepackage{amsmath}
\usepackage{amssymb}
\usepackage{mathtools}
\usepackage{amsthm}

\usepackage[capitalize,noabbrev]{cleveref}

\theoremstyle{plain}

\theoremstyle{definition}

\theoremstyle{remark}

\usepackage[textsize=tiny]{todonotes}

\icmltitlerunning{Agentic Guard Rails for Federated Computing}

\begin{document}

\twocolumn[
\icmltitle{Can One Safety Loop Guard Them All? \\
           Agentic Guard Rails for Federated Computing}




\begin{icmlauthorlist}
\icmlauthor{Narasimha Raghavan Veeraragavan}{yyy}
\icmlauthor{Jan Franz Nygård}{yyy,comp}
\end{icmlauthorlist}

\icmlaffiliation{yyy}{Division of Cancer Registry of Norway, Norwegian Institute of Public Health, Oslo, Norway}
\icmlaffiliation{comp}{Department of Physics and Technology, The Arctic University of Norway, Tromsø, Norway}

\icmlcorrespondingauthor{Narasimha Raghavan Veeraragavan}{narasimha.raghavan.veeraragavan@fhi.no}
\icmlcorrespondingauthor{Jan Franz Nygård}{jan.franz.nygard@fhi.no}

\icmlkeywords{Machine Learning, ICML}

\vskip 0.3in
]



\printAffiliationsAndNotice{} 

\begin{abstract}
We propose Guardian-FC, a novel two-layer framework for privacy preserving federated computing that unifies safety enforcement across diverse privacy preserving mechanisms, including cryptographic back-ends like fully homomorphic encryption (FHE) and multiparty computation (MPC), as well as statistical techniques such as differential privacy (DP). Guardian-FC decouples guard-rails from privacy mechanisms by executing plug-ins (modular computation units), written in a backend-neutral, domain-specific language (DSL) designed specifically for federated computing workflows and interchangeable Execution Providers (EPs), which implement DSL operations for various privacy back-ends. An Agentic-AI control plane enforces a finite-state safety loop through signed telemetry and commands, ensuring consistent risk management and auditability. The manifest-centric design supports fail-fast job admission and seamless extensibility to new privacy back-ends. We present qualitative scenarios illustrating backend-agnostic safety and a formal model foundation for verification. Finally, we outline a research agenda inviting the community to advance adaptive guard-rail tuning, multi-backend composition,  DSL specification development, implementation, and compiler extensibility alongside human-override usability.
\end{abstract}

\section{Introduction}\label{sec:introduction}
Federated computing enables collaborative data analysis and model training across multiple institutions without sharing raw data, offering significant privacy advantages. However, ensuring rigorous privacy guarantees across diverse privacy preserving technologies remains a major challenge. Techniques such as fully homomorphic encryption (FHE), differential privacy (DP), and multiparty computation (MPC) each provide distinct mechanisms to protect data, but they come with bespoke safety checks and enforcement mechanisms that complicate unified governance. 

Each privacy mechanism introduces unique operational constraints and safety metrics:  For example, CKKS pipelines that implement FHE track \emph{noise budget}~\cite{cheon2017ckks},
DP pipelines track \(\varepsilon\) budget~\cite{mironov2017renyi}, MPC pipelines track share integrity~\cite{gamiz2025challenges}.  These differences have led to fragmented safety frameworks, where each privacy back-end requires specialized monitoring, control logic, and auditing tools. 

This fragmentation increases development complexity by requiring separate teams or toolchains to build, maintain, and verify distinct safety mechanisms for each privacy technology, leading to duplicated effort such as implementing multiple versions of monitoring systems, control logic, auditing tools, and safety predicates tailored to each privacy back-end. It hinders interoperability because workflows combining multiple privacy back-ends lack a common safety framework, making it difficult to coordinate policies and share safety guarantees across components. Furthermore, inconsistent enforcement of safety policies across heterogeneous systems raises the risk of privacy breaches or system failures, as gaps or mismatches in monitoring and control can allow unsafe states to go undetected or unmitigated.

Moreover, federated computing workflows often combine multiple privacy techniques. For example, in computing a federated Kaplan–Meier survival curve~\cite{veeraragavan2024multipartyhomomorphicencryptionapproach}, each participating site first encrypts its local survival data, such as event times and censoring indicators, using FHE, which allows computations on encrypted data without revealing raw inputs. The encrypted partial aggregates, like counts of events and censored observations at each time point, are then securely combined across sites using MPC protocols, ensuring correct aggregation without exposing individual data shares. After aggregation, the combined encrypted statistics are jointly decrypted by the parties through MPC, revealing the plaintext aggregate while preserving data confidentiality. To protect against inference attacks on the released survival estimates, calibrated noise is added to the decrypted aggregate following differential privacy (DP) principles, providing formal privacy guarantees. Finally, the noise-perturbed survival estimates are released, enabling accurate yet privacy-preserving survival analysis across distributed datasets. This combination leverages the strengths of FHE for local data confidentiality, MPC for secure aggregation and decryption, and DP for rigorous privacy protection of the published results.

Such multi-stage pipelines currently lack comprehensive end-to-end safety verification, making it difficult to reason about overall privacy guarantees and system health. The absence of a unified safety control plane that can operate seamlessly across heterogeneous privacy back-ends limits scalability and trustworthiness in practical deployments.  

To address these challenges, a reusable, privacy backend agnostic safety framework is essential, one that can sense, predict, and act upon safety risks uniformly, regardless of the underlying privacy mechanism. This framework must also support extensibility to emerging privacy technologies and provide auditable guarantees to operators and regulators. 

To this end, we propose GUARDIAN-FC, a novel two layer framework that unifies safety enforcement across heterogeneous privacy back-ends. \gc decouples guard-rails (the runtime safety policies) from the underlying privacy mechanisms, enabling a reusable backend agnostic safety loop. At its core, federated computations are expressed as plug-ins modular units written in a backend-neutral, domain specific language (DSL) designed specifically for federated computation workflows. These plug-ins are executed by interchangeable Execution Providers (EPs), which implement the DSL operations for various privacy-preserving back-ends such as FHE, DP, and MPC. 

Above the federation layer, an Agentic-AI control plane orchestrates a finite-state safety loop that continuously senses system health via signed telemetry, predicts risks, acts through signed control commands, and proves compliance via an auditable ledger. This design ensures consistent risk management and auditability regardless of the privacy technology in use.

A manifest-centric registration protocol serves as a critical gatekeeper that verifies the compatibility and correctness of federated computing jobs before execution. Each job submission requires, a JSON document that specifies the plugin, the chosen Privacy Execution Provdier (PEP), required DSL operations, and enabled guard-rail predicates. This manifest act as an authoritative contract consumed by all system components, ensuring that every node and the central aggregator share a consistent understanding of the job's configuration without exposing private details. By performing comprehensive admission checks against this manifest, system can fail fast immediately rejecting jobs with incompatible or unsupported configurations, thus preventing wasted computations and potential safety violations. 

Furthermore, this design enables seamless extensibility, allowing new privacy back-ends to be integrated simply by creating new Execution Provider modules that implement the required DSL operations and metrics, without requiring changes to the core control plane or guard-rail logic. This modular and declarative approach streamlines deployment, enhances robustness, and supports evolving privacy technologies within a unified safety framework.

We demonstrate GUARDIAN-FC’s backend-agnostic safety through qualitative scenarios covering noise exhaustion, privacy budget overflow, and malformed shares across different privacy mechanisms. Additionally, we provide a formal finite-state model that lays the foundation for future verification efforts. Finally, we outline a research agenda inviting the community to advance adaptive guard-rail tuning, multi-backend composition, DSL specification and implementation, compiler extensibility, and human-override usability.

By providing a unified, formally grounded safety framework, GUARDIAN-FC aims to accelerate the adoption of privacy-preserving federated computing in real-world applications.

\section{Architecture}
\label{sec:arch}
\begin{figure}[t]
  \centering
  \includegraphics[width=1.1\columnwidth]{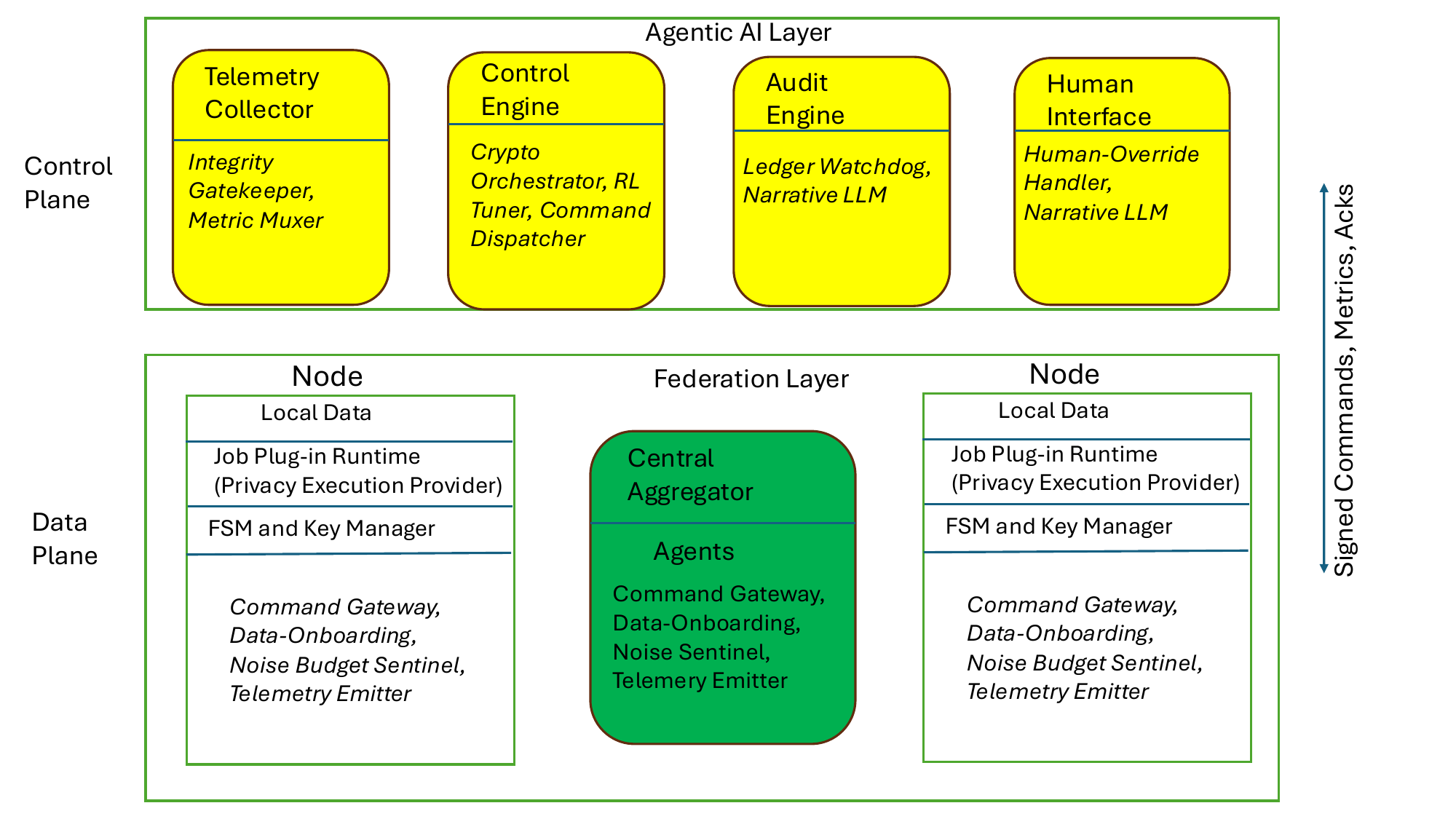}
\caption{Guardian-FC architecture.  
\textbf{Top:} the \emph{Agentic-AI layer}  constitutes the control plane and holds only signed metadata.  
\textbf{Bottom:} the \emph{Federation layer}  constitutes the data plane, where Nodes and a Central Aggregator execute privacy-preserving workloads.  
Upward arrows carry signed metric frames; downward arrows carry signed A-commands and acknowledgments.  
}
\label{fig:arch}
  \end{figure}

GUARDIAN-FC is structured as a two-layer framework that cleanly separates the control plane from the data plane as shown in Figure~\ref{fig:arch}, enabling backend-agnostic safety enforcement for federated computing workloads.
\subsection{Control Plane: Agentic-AI safety loop}

The control plane implements an Agentic-AI control system responsible for monitoring, predicting, and acting upon safety risks during federated computations. It processes only small, cryptographically signed metadata never raw data or ciphertexts to maintain strict confidentiality boundaries. Key components include:

\begin{itemize}[leftmargin=*]
   \item \textbf{Telemetry Collector:}  
        Aggregates cryptographically signed metric frames at a fixed cadence such as \(1\,\mathrm{Hz}\)~(refer Appendix~\ref{app:telemetry}) from all participating Nodes and the Central Aggregator, ensuring data integrity and freshness.   
        
  \item \textbf{Sentinel and Control Engine:}  
        Continuously evaluate safety predicates over aligned telemetry, forecasting risks and triggering guard-rail commands when thresholds are breached.
 \item \textbf{Crypto Orchestrator and Command Dispatcher:}
 Issue signed control commands (A-commands) to Nodes and Aggregator, enforcing safety actions such as job aborts or noise resets.
   \item \textbf{Audit Engine}  
        Maintains an append-only Merkle ledger of all commands and acknowledgments, supporting transparent, tamper-evident compliance auditing.
   \item  \textbf{Feedback Interface}
        Provides operators with a dashboard moderated by language models, enabling human overrides through signed commands while preserving safety guarantees.
\end{itemize}
\subsection{Data Plane: Federated Computation Layer}
The data plane executes the privacy-preserving federated workloads. It consists of multiple Nodes (one per participant) and a Central Aggregator, each running a finite-state machine (FSM) to manage job lifecycle states. Design principles include:
\begin{itemize}[leftmargin=*]
\item \textbf{Local-First Confidentiality:} Raw data never leaves the local Node; only encrypted or secret-shared data is communicated.
\item \textbf{Plug-in Runtime:}  Executes federated analytics plug-ins written in a backend-neutral domain-specific language (DSL). Plug-ins dynamically bind to an interchangeable Execution Provider (EP) implementing the privacy back-end (e.g., FHE, DP, MPC).
\item \textbf{Minimal Trusted Surface:} Each Node and the central aggregator runs exactly one instance of Command Gateway and one Telemetry Emitter to authenticate commands and forward to the FSM and emit signed metrics.
\item \textbf{Symmetric Observability} All participants emit identical metric frames at fixed intervals, enabling consistent global safety evaluation.
\end{itemize}

\subsection{Cross-layer Interface}
Communication between the control and data planes occurs via authenticated inter-process channels:
\begin{itemize}[leftmargin=*]
\item \textbf{Upstream Metrics:} Nodes and Aggregator send signed telemetry frames containing metrics such as noise budget, privacy loss, and latency at 1 Hz.

\item \textbf{Downstream Commands:} The control plane sends idempotent, signed A-commands to Nodes and Aggregator, each requiring acknowledgment before the next telemetry tick.

This architecture enables GUARDIAN-FC to apply a uniform finite-state safety loop across heterogeneous privacy back-ends without modifying plug-in code or control logic.
\end{itemize}

\section{Domain-Specific Language (DSL)}
At the core of GUARDIAN-FC's flexible and backend-agnostic design is a domain specific language (DSL) tailored specfically for federated computing workflows. The DSL enables data scientists and developers to write plug-ins, modular computation units that experess federated computing logic without embedding privacy mechanism details. 

The DSL is designed to be:
\begin{itemize}[leftmargin=*]
\item \textbf{Backend-neutral:} Abstracts away the underlying privacy-preserving technology, allowing the same plug-in code to run on different Execution Providers (EPs) such as FHE, DP, and MPC. 

\item \textbf{Expressive: } Supports the operations and data types commonly required for federated computing, including handling encrypted or secret shared data. 

\item \textbf{Composable:} Enables building complex workflows by combining simpler operations, facilitating reuse and modularity. 
\end{itemize}
\subsection{Plug-ins and Execution Providers}
Plug-ins are written in the DSL and define \textit{what} computation to perform within the Nodes and the Central Aggregator. Execution Providers (EPs) implement the DSL's operational semantics for specific privacy back-ends, defining how computations are executed securely and privately. This separation allows \gc to enforce uniform safety guard-rails regardless of the chosen privacy mechanism, enhancing modularity and extensibility. 
\subsection{Compilation and Runtime}
The compilation process translates high-level DSL code into binaries or modules tailored for particular privacy-preserving technologies, such as FHE, DP, or MPC. At runtime, these compiled plug-in components do not operate in isolation; instead, they bind dynamically to EP modules that implement the concrete semantics of the chosen privacy back-end. This dynamic binding means that plug-ins' computation logic remains unchanged while the EP module provides the specific cryptographic or statistical operations required for privacy enforcement. 

This architecture allows GUARDIAN-FC to seamlessly switch between different privacy back-ends by simply swapping the EP module without recompiling or modifying the plug-in source code. For example, the same plug-in can run under an FHE-based EP for encrypted computation or a DP-based EP for noise addition, depending on the deployment context.
Furthermore, this modular approach supports flexible deployment strategies, such as packaging the plug-in and EP binaries into Docker containers for Nodes and the Central Aggregator. It also facilitates extensibility, enabling privacy engineers to develop new EP modules independently and integrate them without disrupting existing plug-ins or control logic.
Overall, this compilation and runtime design promotes modularity, extensibility, and backend-agnosticism, simplifying development and enhancing the robustness of privacy-preserving federated computing workflows.

\section{Registration Flow}
The registration flow is a critical step in GUARDIAN-FC that governs the admission and setup of federated computing jobs, ensuring that all components, plug-ins, EPs, and guard-rail predicates are mutually compatible before any sensitive data or ciphertext is exchanged. 

At the heart of this process is a manifest-centric protocol. When a job is submitted, a manifest, a structured JSON document, is generated during plug-in compilation and decoration. This manifest specifies key information including the computing plug-in name, the required DSL operations, the chosen EP (such as FHE, DP, or MPC), the guard-rail predicates enabled for the job, and the list of metric keys (such as noise budgets) produced by the EP and emitted by the Nodes and Central Aggregator during execution. This manifest acts as an authoritative contract that is consumed by all system components, including Nodes, the Central Aggregator, and the Agentic-AI control plane. By sharing a consistent configuration without exposing private details, the manifest ensures that every participant operates under the same assumptions.

Before execution begins, the Control Engine performs fail-fast admission checks based on the manifest. It verifies that every DSL opcode used by the plug-in is implemented by the selected EP and confirms that all enabled guard-rail predicates reference metric keys provided by the EP. If any compatibility check fails, the job is immediately rejected, preventing wasted computation and potential safety violations.
Once admitted, the manifest is broadcast to all Nodes and the Central Aggregator. Each Node dynamically loads and binds the specified EP module at runtime, enabling the same plug-in code to run unchanged across different privacy back-ends by simply swapping the EP. This dynamic binding leverages the backend-neutral nature of the DSL and the modularity of EPs introduced earlier.

To maintain strict safety guarantees during execution, the Control Engine publishes the manifest hash and the final list of metric keys to the Telemetry Collector. This component rejects any telemetry frame whose schema or signature deviates from the manifest contract, ensuring that runtime guard-rail evaluation requires no reflection or dynamic schema inference. This guarantees that all telemetry data used for safety monitoring is consistent and trustworthy.
In summary, the registration flow provides a robust and extensible mechanism to ensure safe, consistent, and backend-agnostic job execution in GUARDIAN-FC. It tightly integrates with the DSL and EP architecture, enabling seamless extensibility to new privacy back-ends without modifying core control logic or plug-in source code.

\section{Guard-Rail Predicates}

A \emph{predicate} \(p_i\colon\{\mathbf m,\widehat{\mathbf m}\}
\!\to\!\{\textsc{true},\textsc{false}\}\) is a Boolean expression over
the current metric frame \(\mathbf m\) or the one-tick forecast
\(\widehat{\mathbf m}\) emitted by the Sentinel. These predicates form the core safety checks within GUARDIAN-FC’s control plane, continuously monitoring the health and privacy guarantees of federated computations.
Guard-rail predicates operate on telemetry metrics emitted by the Nodes and the Central Aggregator during execution. By defining clear safety conditions, they enable timely detection of risks such as noise budget exhaustion in fully homomorphic encryption (FHE), privacy budget overflow in differential privacy (DP), or malformed data shares in multiparty computation (MPC).
Example guard-rail rules are listed in Table~\ref{tab:pred-examples}, while the full set of predicates is maintained in a versioned, declarative configuration file named \texttt{guardrails.yaml}, which is loaded at job admission time. Only predicates referencing metric keys provided by the selected Execution Provider (EP) are enabled for a given job, ensuring relevance and correctness.
When a predicate evaluates to true, the control plane issues a corresponding signed control command (A-command) to initiate corrective actions, such as aborting the job, resetting noise budgets, or isolating faulty participants. This standardized approach enforces uniform safety policies across diverse privacy back-ends, simplifying operator oversight and auditability.
In summary, guard-rail predicates provide a formal, backend-agnostic mechanism to sense and respond to safety risks in federated computing, forming a critical component of GUARDIAN-FC’s unified safety framework.

\begin{table}[h]
\small\centering
\begin{tabular}{@{}llc@{}}
\toprule
Predicate & Expression & A-command \\ \midrule
\(p_1\) noise       & \(\mathbf m.\text{noiseBits}<\theta_{\text{fhe}}\) & \texttt{A-BOOTSTRAP}\\
\(p_2\) DP budget   & \(\mathbf m.\varepsilon_{\text{spent}}>
                       \varepsilon_{\max}\) & \texttt{A-ABORT\_JOB}\\
\(p_3\) quorum      & \(\mathbf m.\text{shareAuthFail}\ge 2\) &
                       \texttt{A-ISOLATE\_PARTY}\\ \bottomrule
\end{tabular}
\caption{Illustrative guard-rail predicates.  All share the same type
\(\{\mathbf m,\widehat{\mathbf m}\}\!\rightarrow\!\{\textsc{true},\textsc{false}\}\).}
\label{tab:pred-examples}
\end{table}

\section{Finite-State Safety Loop}

\noindent \textbf{GUARDIAN-FC} enforces safety through a formally specified \emph{finite-state safety loop} that operates continuously during federated computations.  
At a fixed cadence (e.g.\ 1 Hz) the loop executes a recurring
\[
\text{Sense} \;\to\; \text{Predict} \;\to\; \text{Act} \;\to\; \text{Prove}
\]
cycle to monitor privacy guarantees and system health.

\begin{description}
  \item[Sense:] The \emph{Telemetry Collector} gathers \emph{signed metric frames} from all Nodes and from the Central Aggregator.  
  Each frame reports key metrics such as noise budget, privacy loss and latency, and is authenticated to ensure integrity and freshness.
  
  \item[Predict:] The \emph{Sentinel} aligns the telemetry streams and evaluates the enabled \emph{guard-rail predicates} on both current and predicted metrics, forecasting imminent safety violations.

  \item[Act:] When a predicate would be violated, the \emph{Control Engine} instructs the \emph{Crypto Orchestrator} to emit signed \emph{A-commands} to Nodes and to the Aggregator.  
  Actions include aborting jobs, resetting noise budgets, or isolating faulty participants.

  \item[Prove:] The \emph{Audit Engine} appends all commands, acknowledgments and metric hashes to an append-only \emph{Merkle ledger}, creating tamper-evident evidence for compliance reporting and post-hoc analysis.
\end{description}

The loop is modeled as the synchronous product of finite-state machines (FSMs) for
\emph{Nodes}, the \emph{Central Aggregator}, the \emph{Control Engine}, the \emph{Telemetry Collector}, and the \emph{Audit Engine}.  
The combined state-space is kept small enough to enable model checking, so that all safety invariants can be proven and the system is guaranteed to progress to a safe termination state. For a detailed overview of the finite-state machines and their states, please refer to Table~\ref{tab:fsm_states} in Appendix A.

\subsection{Safety Property}
Let
$P=\{p_{1},p_{2},\dots,p_{k}\}$
be the set of enabled guard-rail predicates for the current job, and let
\[
S_{ok}\subseteq\{\texttt{IDLE},\texttt{PREF},\texttt{INF},\texttt{POSTF},\texttt{DONE},\texttt{ABORTED}\} \] denote the set of Node states considered safe at release time 
\[
\quad(\;S_{ok}=\{\texttt{POSTF}\}\text{ by default}\,).
\]

In the safety invariant, the set of Node states considered safe at release time, 
 $S_{ok}$, is by default set to POSTF, representing the post-flight stage. This choice reflects the critical point in the job lifecycle where the core federated computation has completed successfully, and the Node is performing necessary post-processing before finalizing results.
The POSTF state is the minimal safe checkpoint ensuring that all privacy and integrity guarantees hold before the system transitions to finalization. 

The \emph{safety invariant} is
\begin{equation}\label{eq:safety}
\begin{aligned}
\text{Aggregator} &= \texttt{FINALIZE} \;\Longrightarrow\\
&\bigl(\forall i:\text{Node}[i]\in S_{ok}\bigr)
\;\land\;
\bigl(\forall p\in P:\neg p\bigr).
\end{aligned}
\end{equation}

This means that if the Aggregator reaches the \texttt{FINALIZE} state (i.e., the job completes successfully), then every Node must be in a safe state within $S_{ok}$, and no guard-rail predicate is violated. This invariant guarantees that the system only finalizes results when all safety conditions hold, preventing unsafe releases.

\subsection{Liveness Property}
In the liveness property, the ranking function $\mu$ serves as a quantitative measure of the system's progress toward  termination. We define a ranking function $\mu$ over the global system state as the sum of individual ranks assigned to each Node's state and the Aggregator's state and it is expressed as defined in Equation~\ref{eq:rank}: 
\begin{equation}\label{eq:rank}
\mu \;=\;
\sum_{i=1}^{N} \operatorname{rank}\!\bigl(\text{Node}[i]\bigr)
\;+\;
\operatorname{rank}(\text{Aggregator}),
\end{equation}
Each state is assigned with a numeric rank reflecting its distance from a safe terminal conditino, with higher ranks indicating states further from completion.  Accordingly, the following rank values for Node states:
\[
\begin{aligned}
\operatorname{rank}(\texttt{IDLE})   &= 3,\\
\operatorname{rank}(\texttt{PREF})   &= 2,\\
\operatorname{rank}(\texttt{INF})    &= 1,\\
\operatorname{rank}(\texttt{POSTF})  &=
\operatorname{rank}(\texttt{DONE})  =
\operatorname{rank}(\texttt{ABORTED})=0.
\end{aligned}
\]

The purpose of computing $\mu$ is to formally demonstrate that the system cannot remain indefinitely in intermediate sates. As the system executes, $\mu$ is expected to decrease monotonically or remain stable, indicating progress toward termination. When $\mu$ reaches 0, it signifies that all Nodes and the Aggregator have reached safe terminal states, and the job has either completed successfully or aborted safely. 
The \emph{liveness invariant} is expressed as:
\begin{equation}\label{eq:liveness}
\begin{aligned}
\Diamond(\mu=0)\;\equiv\;&\;
\Diamond\!\Bigl(\forall i:\text{Node}[i]\in S_{ok}\Bigr)\\
&{}\land\;
\Diamond\!\Bigl(\text{Aggregator}\in\{\texttt{FINALIZE},\texttt{ABORTED}\}\Bigr).
\end{aligned}
\end{equation}

The computational tree logic (CTL) based Equation~\ref{eq:liveness} expresses that, eventually, every Node reaches a safe state in~$S_{ok}$ and the Aggregator reaches either \texttt{FINALIZE} or \texttt{ABORTED}.  
Hence, the system cannot linger indefinitely in intermediate states; it must always make progress toward a terminal condition.

The safety property in Equation~\eqref{eq:safety} prevents unsafe outcomes, while the liveness property in Equation~\eqref{eq:liveness} guarantees eventual completion or abort of every federated job.  
By promoting the safety loop to a first-class control-plane component with formally verified FSMs, GUARDIAN-FC delivers backend-agnostic, provably correct runtime enforcement of privacy and integrity guarantees across heterogeneous federated-computing environments. 

\section{Qualitative Evaluation}
To illustrate the practical effectiveness of GUARDIAN-FC's backend-agnostic finite-state safety loop, we outline three representative scenarios that show how the system detects and mitigates distinct risk vectors across different privacy-preserving back-ends. Table~\ref{tab:pred-examples} shows the illustrative examples of predicates and corresponding control commands. 

\begin{description}
\item[Scenario A: CKKS Noise Exhaustion]\hfill\\
In a fully homomorphic encryption (FHE) setting using the CKKS scheme, each \emph{Node} tracks a \emph{noise-budget} metric.  
When a Node’s noise budget falls below a configured threshold, the guard-rail predicate $p_{1}$ evaluates to \emph{true}.  
The Control Engine responds by issuing a signed \texttt{A-BOOTSTRAP} command to all Nodes, instructing them to reset their local noise budgets.  
After the acknowledgments arrive, subsequent telemetry frames confirm that the budgets are restored, allowing the computation to continue without a global abort.  
This scenario exemplifies \emph{local repair and continuity} under noise constraints.

\item[Scenario B: Differential-Privacy Budget Overflow]\hfill\\
In a differential-privacy (DP) context, Nodes monitor cumulative privacy loss $\epsilon$.  
If any Node’s $\epsilon$ exceeds the policy limit, predicate $p_{2}$ fires, triggering an \texttt{A-ABORT\;JOB} command.  
All Nodes then securely wipe their DP seeds, and the Central Aggregator discards any pending shares.  
The audit ledger records the abort event, enforcing deterministic \emph{fail-fast} behavior that prevents privacy-budget violations and preserves system integrity.

\item[Scenario C: Malformed MPC Share]\hfill\\
Within a multiparty computation (MPC) framework, the Central Aggregator verifies the zero-knowledge proofs that accompany partial shares.  
Detection of an invalid proof from a Node activates predicate $p_{3}$, prompting an \texttt{A-ISOLATE\;PARTY} command.  
The faulty Node is quarantined, while the Aggregator merges the remaining shares (meeting the quorum threshold) and finalizes the job successfully.  
No changes to the control-plane logic are required, demonstrating graceful \emph{fault isolation}.
\end{description}
In all three cases, the \emph{core safety-loop logic}, component boundaries, and audit semantics remain unchanged.  
Only the fired predicate and the corresponding \texttt{A-command} differ, highlighting GUARDIAN-FC's ability to enforce uniform safety policies across heterogeneous privacy back-ends.  
These examples demonstrate the design’s modularity, extensibility, and robustness in mitigating diverse operational risks.
\section{Open Research Directions}

While GUARDIAN-FC establishes a foundational \emph{backend-agnostic} safety loop for federated computing, several open questions remain that affect adaptability, composability, usability, and extensibility.  
We highlight four promising avenues for future work:

\begin{enumerate}[label=\textbf{\arabic*.},leftmargin=*]

\item \textbf{Adaptive Guard-Rail Policy Tuning}

Static thresholds such as the initial CKKS noise floor
$\theta^{\text{init}}_{\mathrm{fhe}}$
and the initial differential-privacy budget
$\epsilon^{\max}_{\text{init}}$
can be either overly conservative or too lax for a given workload.  
We hypothesise that \emph{reinforcement learning} (RL) techniques could dynamically adjust these thresholds \emph{per job} while still respecting safety invariants.  
A \emph{shielded} RL agent would propose updates that pass through symbolic safety predicates, guaranteeing that no unsafe configuration is deployed.  
Such adaptivity promises better resource utilisation and fewer unnecessary aborts.

\item \textbf{Multi-Execution-Provider Composition}

Real-world workflows often chain heterogeneous privacy mechanisms
(e.g., FHE $\!\to$ DP $\!\to$ MPC), yet current validators check each stage in
isolation, so end-to-end privacy and entropy guarantees remain opaque.
We introduce a \emph{typed} $(\epsilon,\delta,\lambda)$-calculus~\cite{church1940formulation} that
composes differential-privacy budgets $\epsilon,\delta$ together with
cryptographic entropy $\lambda$ (e.g., min-entropy of keys) across
execution-provider boundaries~\cite{reed2010distance}.
Each stage’s type is refined by an Satisfiability Modulo Theories (SMT)-checkable predicate~\cite{de2011satisfiability}; A job is admitted only if the composed budget satisfies \(\varepsilon_{\text{tot}}\le\varepsilon_{\max}\), \(\delta_{\text{tot}}\le\delta_{\max}\), and \(\lambda_{\text{entropy}}\ge\lambda_{\min}\), as specified in a policy file. The deliverable would be a static checker that either accepts any FHE \(\to\) DP \(\to\) MPC pipeline meeting these constraints, or else returns a counterexample (specifying the violating stage and metric) at admission time. A key challenge is reconciling continuous DP loss \((\varepsilon,\delta)\) with discrete key-entropy \(\lambda\) in a single lattice that can be handled by SMT-based reasoning~\cite{barrett2021satisfiability}.

\item \textbf{Human-Override UX and Alert Fatigue}

Excessive or poorly designed alerts can desensitise operators, leading to unsafe overrides.  
We plan to study alert modalities (e.g.\ email, sms, chat messages) together with friction mechanisms (e.g.\ mandatory signatures) to minimise false-positive rates while preserving timely intervention.  
Controlled user studies measuring false positives, response times, and cognitive load (e.g.\ NASA-TLX scores~\cite{hart1988development}) will inform UX guidelines and predictive models.  
Balancing rapid response against operator fatigue is critical for safe human-in-the-loop control.

\item \textbf{Domain-Specific Language (DSL) Specification \& Implementation}

A backend-neutral DSL powers compute plug-ins that bind dynamically to EPs.  
Enhancing its expressiveness, type safety, and extensibility is vital for sophisticated analytics.  
Research tasks include formalising syntax/semantics, building robust compiler tool-chains, and integrating static analyses that verify compatibility with guard-rail predicates and privacy budgets.  
Efficient runtimes capable of hot-binding new EPs will foster extensibility to emerging privacy back-ends, boosting developer productivity and overall system reliability.

\end{enumerate}
\section{Related Work}
CKKS noise-tracking monitors in
HEAAN \cite{cheon2017ckks} and OpenFHE \cite{al2022openfhe}
instrument ciphertext depth but embed policy checks directly in job
code.  Multi-party FHE schemes such as TFHE-MPC \cite{riazi2018chameleon}
combine partial decryption at the server without a separate control
plane.  By contrast, Guardian-FC pushes all policy evaluation into an
orthogonal Agentic-AI layer, allowing the same job plug-in to run under
FHE, DP, or MPC.

Rényi accountants
\cite{mironov2017renyi}, Opacus for PyTorch~\cite{yousefpour2021opacus}, and TensorFlow-Privacy
\cite{papernot2019machine} track composite $\varepsilon$ budgets during stochastic gradient descent (SGD);
however, they provide no cross-backend safety loop.  Our Execution-
Provider API treats the accountant as a drop-in module supervised by the
same guard-rail predicates used for FHE.

 Early actively secure MPC protocols such as
SPDZ~\cite{dpsz2012} introduced
\emph{MAC-authenticated shares}: each party holds a share plus a one-time
information-theoretic MAC; during online evaluation any malformed share
is detected when MACs fail to open.  Subsequent work optimised MAC
length and preprocessing cost (e.g.\ TinyOT/TinyKeys
\cite{hazay2018tinykeys}) and integrated the idea into productions
frameworks such as MP-SPDZ~\cite{keller2020mpspdz}.  A complementary
line of work adds zero-knowledge proofs at the merge point so that an
honest-majority Aggregator can reject invalid ciphertexts without
learning the underlying secret; a recent example include Masquerade’s
verifiable commitments~\cite{mouris2025masquerade}.  
Existing reviews~\cite{evans2018pragmatic, sun2025sok, gamiz2025challenges} agree that today’s MPC frameworks detect a faulty share
but leave \emph{isolation} or \emph{operator response} to out-of-band,
manual procedures.  Guardian-FC closes that gap by lifting isolation
into a signed control-plane command (\texttt{A-ISOLATE\_PARTY}) that is
audited and acknowledged within the same finite-state loop. 

Safety work on large language models (LLMs) has converged on a pattern
in which a \emph{secondary filter} checks every response before it
reaches the user.  \emph{Sparks of AGI}~\cite{bubeck2023sparks}
hand-codes GPT-4 prompts to block tool misuse, while
Self-CheckGPT~\cite{manakul2023selfcheckgpt} queries a second LLM to
detect hallucinations.  The open-source
\textsc{Guardrails-AI} library~\cite{guardrailsai2025docs} lets
developers declare JSON or regex constraints that an LLM output must
satisfy prior to release.  Guardian-FC adopts the same
\emph{Sense→Predict→Act} loop but grounds it in finite-state machines
and cryptographic signatures, targeting privacy-preserving
\emph{computation} rather than text generation.

\section{Conclusion}
GUARDIAN-FC introduces a formally verifiable, finite-state safety loop atop existing FHE, DP, and MPC back-ends, enforcing uniform guard-rails through a backend-neutral DSL and signed telemetry/command channels.  
This decoupled control plane delivers auditable, runtime protection without altering the underlying privacy mechanisms.  
We invite the community to explore \emph{integrating \gc ideas with existing federated computing frameworks}, so that its “one loop to guard them all” can harden real-world deployments of privacy-preserving federated computing.

\nocite{langley00}

\bibliography{example_paper}

\begin{thebibliography}{22}
\providecommand{\natexlab}[1]{#1}
\providecommand{\url}[1]{\texttt{#1}}
\expandafter\ifx\csname urlstyle\endcsname\relax
  \providecommand{\doi}[1]{doi: #1}\else
  \providecommand{\doi}{doi: \begingroup \urlstyle{rm}\Url}\fi

\bibitem[AI()]{guardrailsai2025docs}
AI, G.
\newblock Guardrails ai documentation.
\newblock \url{https://www.guardrailsai.com/docs}.
\newblock Accessed 2025-05-19.

\bibitem[Al~Badawi et~al.(2022)Al~Badawi, Bates, Bergamaschi, Cousins, Erabelli, Genise, Halevi, Hunt, Kim, Lee, et~al.]{al2022openfhe}
Al~Badawi, A., Bates, J., Bergamaschi, F., Cousins, D.~B., Erabelli, S., Genise, N., Halevi, S., Hunt, H., Kim, A., Lee, Y., et~al.
\newblock Openfhe: Open-source fully homomorphic encryption library.
\newblock In \emph{proceedings of the 10th workshop on encrypted computing \& applied homomorphic cryptography}, pp.\  53--63, 2022.

\bibitem[Barrett et~al.(2021)Barrett, Sebastiani, Seshia, and Tinelli]{barrett2021satisfiability}
Barrett, C., Sebastiani, R., Seshia, S.~A., and Tinelli, C.
\newblock Satisfiability modulo theories.
\newblock In \emph{Handbook of satisfiability}, pp.\  1267--1329. IOS Press, 2021.

\bibitem[Bubeck et~al.(2023)Bubeck, Chandrasekaran, Eldan, et~al.]{bubeck2023sparks}
Bubeck, S., Chandrasekaran, V., Eldan, R., et~al.
\newblock Sparks of artificial general intelligence: Early experiments with {GPT‐4}.
\newblock \emph{arXiv preprint}, arXiv:2303.12712, 2023.

\bibitem[Cheon et~al.(2017)Cheon, Kim, Kim, and Song]{cheon2017ckks}
Cheon, J.~H., Kim, A., Kim, M., and Song, Y.
\newblock Homomorphic encryption for arithmetic of approximate numbers.
\newblock In \emph{Advances in cryptology--ASIACRYPT 2017: 23rd international conference on the theory and applications of cryptology and information security, Hong kong, China, December 3-7, 2017, proceedings, part i 23}, pp.\  409--437. Springer, 2017.

\bibitem[Church(1940)]{church1940formulation}
Church, A.
\newblock A formulation of the simple theory of types.
\newblock \emph{The journal of symbolic logic}, 5\penalty0 (2):\penalty0 56--68, 1940.

\bibitem[Damg{\aa}rd et~al.(2012)Damg{\aa}rd, Pastro, Smart, and Zakarias]{dpsz2012}
Damg{\aa}rd, I., Pastro, V., Smart, N.~P., and Zakarias, S.
\newblock Multiparty computation from somewhat homomorphic encryption.
\newblock In \emph{CRYPTO 2012}, volume 7417 of \emph{LNCS}, pp.\  643--662, 2012.
\newblock Introduces MAC-authenticated shares that underpin the SPDZ family of protocols.

\bibitem[De~Moura \& Bj{\o}rner(2011)De~Moura and Bj{\o}rner]{de2011satisfiability}
De~Moura, L. and Bj{\o}rner, N.
\newblock Satisfiability modulo theories: introduction and applications.
\newblock \emph{Communications of the ACM}, 54\penalty0 (9):\penalty0 69--77, 2011.

\bibitem[Evans et~al.(2018)Evans, Kolesnikov, Rosulek, et~al.]{evans2018pragmatic}
Evans, D., Kolesnikov, V., Rosulek, M., et~al.
\newblock A pragmatic introduction to secure multi-party computation.
\newblock \emph{Foundations and Trends{\textregistered} in Privacy and Security}, 2\penalty0 (2-3):\penalty0 70--246, 2018.

\bibitem[Gamiz et~al.(2025)Gamiz, Regueiro, Lage, Jacob, and Astorga]{gamiz2025challenges}
Gamiz, I., Regueiro, C., Lage, O., Jacob, E., and Astorga, J.
\newblock Challenges and future research directions in secure multi-party computation for resource-constrained devices and large-scale computations.
\newblock \emph{International Journal of Information Security}, 24\penalty0 (1):\penalty0 1--29, 2025.

\bibitem[Hart \& Staveland(1988)Hart and Staveland]{hart1988development}
Hart, S.~G. and Staveland, L.~E.
\newblock Development of nasa-tlx (task load index): Results of empirical and theoretical research.
\newblock In \emph{Advances in psychology}, volume~52, pp.\  139--183. Elsevier, 1988.

\bibitem[Hazay et~al.(2018)Hazay, Orsini, Scholl, and Soria-Vazquez]{hazay2018tinykeys}
Hazay, C., Orsini, E., Scholl, P., and Soria-Vazquez, E.
\newblock Concretely efficient large-scale {MPC} with active security (\emph{a.k.a.\ TinyKeys for TinyOT}).
\newblock In \emph{ASIACRYPT 2018}, LNCS, 2018.
\newblock Shows short-key information-theoretic MACs for integrity.

\bibitem[Keller(2020)]{keller2020mpspdz}
Keller, M.
\newblock {MP{-}SPDZ}: A versatile framework for multi-party computation.
\newblock In \emph{ACM CCS 2020}, 2020.
\newblock Open-source implementation of SPDZ with MAC-based integrity.

\bibitem[Manakul et~al.(2023)Manakul, Liusie, and Gales]{manakul2023selfcheckgpt}
Manakul, P., Liusie, A., and Gales, M.~J.
\newblock Selfcheckgpt: Zero-resource black-box hallucination detection for generative large language models.
\newblock \emph{arXiv preprint arXiv:2303.08896}, 2023.

\bibitem[Mironov(2017)]{mironov2017renyi}
Mironov, I.
\newblock R{\'e}nyi differential privacy.
\newblock In \emph{2017 IEEE 30th computer security foundations symposium (CSF)}, pp.\  263--275. IEEE, 2017.

\bibitem[Mouris \& Tsoutsos(2025)Mouris and Tsoutsos]{mouris2025masquerade}
Mouris, D. and Tsoutsos, N.~G.
\newblock Masquerade: Verifiable multi-party aggregation with secure multiplicative commitments.
\newblock \emph{ACM Transactions on Internet Technology}, 25\penalty0 (1):\penalty0 1--31, 2025.

\bibitem[Papernot(2019)]{papernot2019machine}
Papernot, N.
\newblock Machine learning at scale with differential privacy in $\{$TensorFlow$\}$.
\newblock In \emph{2019 $\{$USENIX$\}$ Conference on Privacy Engineering Practice and Respect ($\{$PEPR$\}$ 19)}, 2019.

\bibitem[Reed \& Pierce(2010)Reed and Pierce]{reed2010distance}
Reed, J. and Pierce, B.~C.
\newblock Distance makes the types grow stronger: a calculus for differential privacy.
\newblock In \emph{Proceedings of the 15th ACM SIGPLAN international conference on Functional programming}, pp.\  157--168, 2010.

\bibitem[Riazi et~al.(2018)Riazi, Weinert, Tkachenko, Songhori, Schneider, and Koushanfar]{riazi2018chameleon}
Riazi, M.~S., Weinert, C., Tkachenko, O., Songhori, E.~M., Schneider, T., and Koushanfar, F.
\newblock Chameleon: A hybrid secure computation framework for machine learning applications.
\newblock In \emph{Proceedings of the 2018 on Asia conference on computer and communications security}, pp.\  707--721, 2018.

\bibitem[Sun \& Makri(2025)Sun and Makri]{sun2025sok}
Sun, S. and Makri, E.
\newblock Sok: Multiparty computation in preprocessing model.
\newblock \emph{Cryptology ePrint Archive}, 2025.

\bibitem[Veeraragavan et~al.(2024)Veeraragavan, Boudko, and Nygård]{veeraragavan2024multipartyhomomorphicencryptionapproach}
Veeraragavan, N.~R., Boudko, S., and Nygård, J.~F.
\newblock A multiparty homomorphic encryption approach to confidential federated kaplan meier survival analysis, 2024.
\newblock URL \url{https://arxiv.org/abs/2412.20495}.

\bibitem[Yousefpour et~al.(2021)Yousefpour, Shilov, Sablayrolles, Testuggine, Prasad, Malek, Nguyen, Ghosh, Bharadwaj, Zhao, et~al.]{yousefpour2021opacus}
Yousefpour, A., Shilov, I., Sablayrolles, A., Testuggine, D., Prasad, K., Malek, M., Nguyen, J., Ghosh, S., Bharadwaj, A., Zhao, J., et~al.
\newblock Opacus: User-friendly differential privacy library in pytorch.
\newblock \emph{arXiv preprint arXiv:2109.12298}, 2021.

\end{thebibliography}
\bibliographystyle{icml2025}

\newpage
\appendix
\onecolumn
\section{1 Hz Telemetry Frame Schema}\label{app:telemetry}

\begin{table}[h]
\centering
\caption{JSON–protobuf metric frame emitted at 1 Hz by every Node.
Fields absent in a given privacy back-end are set to \texttt{null}.}
\begin{tabular}{@{}llp{4cm}@{}}
\toprule
\textbf{Field} & \textbf{Example (FHE)} & \textbf{Purpose} \\ \midrule
\texttt{node\_id}          & "hospital\_A"               & Stable identifier, also hashed into ledger \\[2pt]
\texttt{seq}               & 17\,429                     & Detects drops / replays \\[2pt]
\texttt{noiseBits}         & 29                          & FHE confidentiality margin \\[2pt]
\texttt{levelsLeft}        & 3                           & Bootstrap trigger threshold \\[2pt]
\texttt{epsilonSpent}      & null                        & DP budget (field is \texttt{0.72} in DP jobs) \\[2pt]
\texttt{lag\_ms}           & 135                         & Straggler / availability metric \\[2pt]
\texttt{opLatency\_ms}     & 2.3                         & Input to RL tuner \\[2pt]
\texttt{timestamp}         & 2025-05-15T10:37:54Z        & Ordering on collector side \\[2pt]
\texttt{sig}               & ed25519:4c8ef…              & Frame authenticity and anti-replay \\ \bottomrule
\end{tabular}

\end{table}

\section{Finite-State Machines and their States}

\begin{table}[h]
\centering
\caption{Finite-State Machines and Their States in \gc}
\begin{tabularx}{\linewidth}{@{}l l L{0.45\linewidth}@{}}
\toprule
\textbf{Component} & \textbf{State} & \textbf{Description} \\
\midrule
\multirow{6}{*}{Node FSM}
  & IDLE     & Waiting for job start \\
  & PREF     & Preparing resources / loading data (pre-flight) \\
  & INF      & Executing federated computation (in-flight) \\
  & POSTF    & Post-processing after computation (post-flight) \\
  & DONE     & Job completed successfully \\
  & ABORTED  & Aborted due to error or safety violation \\
\midrule
\multirow{4}{*}{Aggregator FSM}
  & WAIT     & Awaiting encrypted shares from Nodes \\
  & MERGE    & Securely combining partial shares \\
  & FINALIZE & Producing aggregate result \\
  & ABORTED  & Job aborted \\
\midrule
\multirow{3}{*}{Control Engine FSM}
  & READY    & Idle metrics available for check \\
  & EVALUATE & Checking guard-rail predicates \\
  & DISPATCH & Sending A-commands to participants \\
\midrule
\multirow{3}{*}{Telemetry Collector FSM}
  & OPEN     & Receiving telemetry frames \\
  & WRITE    & Buffering signed metrics \\
  & SEAL     & Closing / sealing current batch \\
\midrule
\multirow{2}{*}{Audit Engine FSM}
  & APPEND   & Adding records to Merkle ledger \\
  & COMMIT   & Committing sealed ledger block \\
\bottomrule
\end{tabularx}
\label{tab:fsm_states}
\end{table}

\end{document}